\documentclass[reprint, amsmath,amssymb, aps, prb]{revtex4-2}

\usepackage{graphicx}
\usepackage{dcolumn}
\usepackage{bm}
\usepackage{xcolor}

\newcommand{\ds}{\displaystyle}

\newcommand{\rev}[1]{{\color{black}#1}}

\begin{document}

\preprint{APS/123-QED}

\title{Non-reciprocal electron transport in finite-size superconductor/ferromagnet bilayers with strong spin-orbit coupling}

\author{A. V. Putilov}
\affiliation{Institute for Physics of Microstructures, Russian Academy of Sciences, 603950 Nizhny Novgorod, GSP-105, Russia}
\affiliation{Moscow Institute of Physics and Technology, Dolgoprudnyi, Moscow Region 141701, Russia}
\email{alputilov@ipmras.ru}
\author{S. V. Mironov}
\affiliation{Institute for Physics of Microstructures, Russian Academy of Sciences, 603950 Nizhny Novgorod, GSP-105, Russia}
\author{A. I. Buzdin}
\affiliation{University Bordeaux, LOMA UMR-CNRS 5798, F-33405 Talence Cedex, France}
\affiliation{World-Class Research Center ``Digital Biodesign and Personalized Healthcare'', Sechenov First Moscow State Medical University, Moscow, 19991, Russia}
\date{\today}

\begin{abstract}
We show that spin-orbit coupling at the interface between a superconducting film of the finite lateral size and the underlying ferromagnetic insulator with in-plane exchange field gives rise to a series of non-reciprocal effects provided the superconducting pairing is enhanced near the boundaries of the superconductor due to, e.g., variation of the film thickness or of the interlayer electron transparency. Specifically, the critical temperature and the critical depairing current are shown to depend on the relative orientation between the exchange field in the ferromagnet and the superconducting film boundaries. The discovered anisotropy of the superconducting properties is promising for the design of diode-type elements in superconducting spintronics.  
\end{abstract}

\maketitle

\section{Introduction}

Interplay between strong spin-orbit coupling (SOC) and ferromagnetic (F) ordering of electron spins in superconducting (S) systems has attracted much attention because it appeared to induce a variety of fascinating phenomena including the appearance of Majorana edge states \cite{Alicea}, Josephson $\varphi_0$ junctions \cite{Buzdin_Phi, Krive, Reynoso, Kouwenhoven}, helical states with spontaneous electric currents \cite{Mironov-PRL-17, Devizorova-PRB-21} etc. 

In systems with the lack of inversion symmetry along a certain direction characterized by the vector ${\bf n}$ the electron free energy contains a term $\propto \left({\bf \sigma}\times{\bf n}\right)\cdot {\bf p}$ which couples the electron momentum ${\bf p}$ with its spin $\sigma$. If such system  is put in electric contact with a ferromagnetic material characterized by the exchange field vector ${\bf h}$ (or is placed into external magnetic field) the electron spins become polarized making the states with two opposite momentum directions along the vector $\left({\bf h}\times{\bf n}\right)$ non-equivalent. As a result, in superconducting media the Cooper pair wave function $\Psi$ in the ground state acquires the spontaneous momentum ${\bf p}_0\propto \left({\bf h}\times{\bf n}\right)$ taking the form of the plane wave $\psi\propto e^{i{\bf p}_0{\bf r}}$ (so-called ``helical'' state) \cite{Edelstein_HelicalStates, Mineev_HelicalStates, Agterberg_review}.  

Despite the comprehensibility of the above physical picture, the consequences of the helical states formation in superconducting systems appeared to be extremely sensitive to the origin of the magnetic ordering and the specific geometry of the system. In particular, for the bulk superconductor placed into the uniform magnetic field an appropriate phase transformation $\Psi^\prime=\Psi e^{-i{\bf p}_0{\bf r}}$ applied to the superconducting order parameter $\Psi$ completely eliminates the effect of SOC on the system free energy so that no meaningful phenomena appears  \cite{Mineev_HelicalStates, Agterberg}. This conclusion is valid also for the bulk ferromagnetic superconductors with the uniform exchange field. However, for systems with non-uniform exchange field there is no phase transformation which can cancel the contribution from the SOC in the whole system, and the helical states appear to reveal themselves in the measurable quantities (e.g., superconducting critical temperature, critical magnetic field, critical depairing current). Specifically, it was shown that the system consisting of a half-infinite bulk superconductor covered by the layer of the ferromagnetic insulator supports the formation of the spontaneous current flowing along the S/F interface which does not affect the critical temperature but gives rise to the stray magnetic fields as well as change the slope of the critical field $H_{c3}$ (corresponding to the emergence of surface superconductivity) as a function of temperature \cite{Mironov-PRL-17}. The decrease of the S layer thickness down to values of the order of the superconducting coherence length $\xi$ (which may be viewed as a decrease in the S sample dimensionality) results in the changes of the superconducting critical temperature and in this case even if the superconducting material belongs to the type-II family the superconducting phase transition may become the first order \cite{Devizorova-PRB-21}. 

In addition, the finite S layer thickness in S/F bilayers with strong SOC is responsible for the appearance of non-reciprocal effects in the superconducting transport, namely, anisotropy of the critical depairing current in the plane of S/F interface \cite{Devizorova-PRB-21}. Choosing the transport current between the minimal and the maximal values of the critical depairing current of the sample one may have zero and finite resistance for two opposite current directions which reminds the diode effect. \rev{Recently the diode effect in superconducting systems has attracted much attention of both theoreticians and experimentalists since it has promising perspectives for application in superconducting electronics and spintronics (see, e.g., Ref.~\cite{Nadeem-NatRP-23} for review). Up to now several classes of superconducting systems simultaneously supporting the electron spin polarization and some sort of SOC were shown to demonstrate the non-reciprocal transport properties. Among them one recognizes different types of hybrid structures where the superconductor is put in electric contact with ferromagnets \cite{Narita-NatNano-22}, topological insulators \cite{Yasuda} or conductive nanowires \cite{Sundaresh-NatComm-23}, [Nb/V/Ta]$_n$ superlattices \cite{Ando-Nature-21}, solid superconducting materials including SrTiO$_3$ \cite{Itahashi, Schumann}, MoS$_2$ \cite{Wakatsuki}, NbSe$_2$ \cite{Bauriedl-NatComm-22}, p-wave WS$_2$ nanotubes \cite{Qin}, Weyl semi-metal T$_d$-MoTe$_2$ \cite{Cui} etc. } From the theoretical standpoint, the non-reciprocal superconducting transport \rev{in systems with the simultaneous spin-polarization and SOC} can be explained, e.g., by accounting the high-order gradients terms in the Ginzburg--Landau theory due to the SOC  \cite{Daido-PRL-22,He-NJP-22} or within microscopic approach \cite{Karabassov-PRB-22,Ilic-PRL-22,Yuan-PNas-22} as well as by considering the Meissner screening of the spontaneous currents generated by the SOC at the S/F interface \cite{Devizorova-PRB-21}. \rev{Interestingly, the diode effect was recently predicted in curved nanowires with SOC and induced superconductivity \cite{Kopasov-PRB-21}. Also a number of theoretical models provides the description for the diode effect in different types of superconductors where the interplay between electron spins and momentum arises due to the peculiarities of the electron band structure instead of Rashba SOC (see, e.g., Refs.~\cite{Scammell, Zinkl, Legg}). Another large class of systems where the diode effect was observed incorporates Josephson junctions with the weak links made of various semiconductors \cite{Baumgartner-NatNano-21, Turini}, ferromagnets \cite{Jeon, Strambini}, Dirac and Weyl semi-metals \cite{Chen, Pal}, graphene \cite{deVries, Diez-Merida} etc. (see also, e.g., Refs.~\cite{Fominov-PRB-22, Davydova, Steiner-PRL-23, Paolucci, Costa} for the relevant theoretical models).} Note also that in a resistive state the asymmetry of the current-voltage characteristics of superconducting systems can be caused, e.g., by non-reciprocal motion of the vortex lattice due to the special form of vortex pinning centers with lack of inversion symmetry \cite{Villegas-Science-03,deSilva-NatLett-06,Gutfreund} or non-uniform distribution of pinning centers \cite{Lyu-NatComm-21}. \rev{Somewhat similar mechanisms responsible for the diode effect were also studied in Refs.~\cite{Carapella, Aladyshkin_1, Aladyshkin_2}.}

In the present paper we show that the finite lateral sizes of the superconducting film placed on top of the ferromagnetic insulator  are responsible for a series of peculiar non-reciprocal effects including the dependence of the superconducting transition critical temperature on the relative orientation between the exchange field in the ferromagnet and the S film boundaries as well as the diode effect in the critical current. We focus on the case when the superconducting transition temperature is locally enhanced near the boundaries of the superconducting film caused by the possible relaxation of the inter-atomic distances at atomic scale or variations of the S layer thickness due to peculiarities of the sample fabrication process.  Similarly to the phenomenon of the twinning-plane superconductivity \cite{Buzdin-JETP-81, Averin-JETP-83, Khlustikov-UFN-88}, the local enhancement of the critical temperature near the S film boundaries gives rise to the nucleation of the localized superconducting states at temperatures above the bulk critical temperature. We show that the phase transition temperature corresponding to the formation of these quasi-one-dimensional states depends on the angle between the exchange field in the underlying ferromagnet and the S film boundary being maximal when the exchange field and the boundary are perpendicular to each other. In addition, when the exchange field is perpendicular to the S layer boundary the critical depairing current of the localized quasi-one-dimensional superconducting channel becomes different for the opposite directions of the transport current producing, thus, the diode-like effect. Such anisotropy is promising for the design of the current controlling elements of superconducting electronics and spintronics. 

Recently, it was experimentally demonstrated that the critical temperature and the critical current of the thin elongated superconducting stripe made of Aluminum and placed on top of dielectric ferromagnet - yttrium iron garnet (YIG) with in-plane magnetization depend on the mutual orientation between the current flow and ferromagnet magnetization \cite{Tikhomirov-JMMM-21}. Although the authors of \cite{Tikhomirov-JMMM-21} consider their results to be caused primary by the effect of stray magnetic fields produced by YIG, our calculation show that similar phenomenon may arise due to the non-reciprocal effects related to the SOC.

The paper is organized as follows. In Sec.~\ref{Sec_Model} we describe the geometry of the S/F system under consideration and introduce the theoretical model based on the Ginzburg-Landau approach. In Sec.~\ref{Sec_Tc} we analyze the dependence of the system critical temperature on the orientation between the exchange field in the F layer and the S layer boundary. In Sec.~{\ref{Sec_Ic}} we describe the diode effect in the critical current. Finally, in Sec.~\ref{Sec_Conc} we summarize our results.

\section{\label{Sec_Model}Theoretical model}

\begin{figure}[b]
\includegraphics[width=0.95\linewidth]{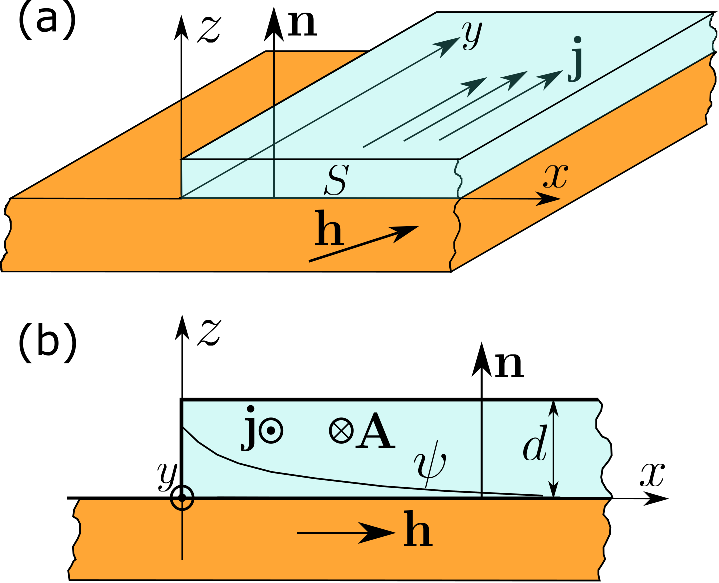}
\caption{Sketch of the boundary of a superconducting film placed on top of ferromagnetic insulator. The exchange field {\bf h} is parallel to the interface between the layers.} 
\label{Fig_1}
\end{figure}

In this section we describe the theoretical approach based on the phenomenological Ginzburg-Landau formalism which we apply to calculate the critical temperature and the critical depairing current of the finite-size S/F system with SOC. The system geometry is sketched in Fig.~\ref{Fig_1}. A superconducting film of the thickness $d$ with the flat boundary is placed on top of the ferromagnetic insulator. A coordinate system is chosen in a way that the superconducting film is parallel to {\it xy} plane with an edge oriented along the {\it y} axis. The {\it z} axis is chosen  perpendicular to the S/F interface and, thus, parallel to the vector $\mathbf n$ characterizing the direction with the broken inversion symmetry far from the S film edge.  So, the S film occupies the region  $x>0,\,0<z<d$ while the F layer occupies the region $z<0$.

In the vicinity of the critical temperature the emergence of superconductivity and the transport properties of the system can be described in the frames of the Ginzburg-Landau theory. For simplicity we neglect the inverse proximity effect assuming that Cooper pairs do not penetrate into the ferromagnetic insulator. Then the system free energy is determined only by the profile of the order parameter inside the superconductor and the free energy density takes the form
\begin{multline}\label{GL_gen}
f= -\left[\alpha(T)+\alpha_1(x)\right]|\psi|^2+\frac{\beta}{2}|\psi|^4+
\varkappa\left|\nabla\psi\right|^2-\\
-\frac{\varepsilon+\varepsilon_1(x)}{2}[\mathbf h_0\times\mathbf n](i\psi^\ast\nabla\psi+c.c.),
\end{multline}
where $\nabla$ is \rev{a gradient operator in} the $xy$ plane, $\alpha(T)=\alpha_0(T_{c0}-T)/T_{c0}$ and $\beta$ are the standard Ginzburg-Landau parameters, $T_{c0}$ is the critical temperature of the isolated bulk superconductor, $\mathbf n$ is the unit vector perpendicular to the S/F interface, $\mathbf h$ is the exchange field directed in the plane of the ferromagnet boundary and the last term proportional to $[\mathbf h_0\times\mathbf n]$ describes the effects of SOC \cite{Samokhin-PRB-04,Kaur-PRL-05,Mironov-PRL-17}. We assume that the S layer thickness in the $z$ direction is much smaller than the superconducting coherence length $\xi$, which allows us to neglect the variation of the order parameter and all coefficients in Eq.~(\ref{GL_gen}) considering all values to be averaged over the film thickness $d$. We assume that the finite lateral size of the superconducting film has a twofold effect on the properties of the S film. First, the strength of the SOC in the vicinity of the S film boundary may substantially differ from the one at $x\gg \xi_0$ (here $\xi_0$ is the zero-temperature coherence length). To account this difference in Eq.~(\ref{GL_gen}) we introduce the value $\varepsilon$ which stands for the spin-orbit coupling constant far from the boundary as well as the function $\varepsilon_1(x)$ which is nonzero only in the region of the width $\sim\xi_0$ near the boundary $x=0$ and describe the difference between the edge SOC constant and $\varepsilon$. Second, the boundary may affect locally the superconducting transition temperature. \rev{This effect can originate from the local variation of the SOC constant since in the presence of the SOC the superconducting critical temperature contains the negative contribution $\propto \varepsilon^2$.} Also the increase in $T_c$ near the sample edge may originate from the local variation of the S layer thickness or the electron transparency of the S/F interface. Assuming the local increase in the critical temperature we introduce the function $\alpha_1(x)$ in Eq.~(\ref{GL_gen}) which is nonzero only at $x\lesssim\xi_0$.

The further simplification can be performed provided the characteristic decay scale of the order parameter in the $x$ direction strongly exceeds the scale $\xi_0$ where the critical temperature and the SOC constant experience the local variations. Then one may put $\alpha_1(x)=\gamma_0\delta(x)$, $\varepsilon_1(x)=\gamma_1\delta(x)$ where $\gamma_0$ and $\gamma_1$ are the certain constants and the delta-function has the spatial scale $\sim\xi_0$. Within the above assumptions we get the following expression for the free energy density:
\begin{equation}
\begin{array}{c}{\ds
\rev{f=-\alpha|\psi|^2+\varkappa|\nabla\psi|^2-\frac{i\gamma_0}{2}\left(\mathbf w\psi^\ast\nabla\psi-c.c.\right)}}\\{}\\{\ds
\rev{-\delta(x)\left[\gamma_0|\psi|^2+\frac{i\varkappa}{2}\left({\bf g}\psi^\ast\nabla\psi-c.c.\right)\right]+\frac{\beta}{2}|\psi|^4,}}
\end{array}\label{Eq_2D}
\end{equation}
where $\mathbf w=\varepsilon[\mathbf h_0\times\mathbf n]/\gamma_0$, and
${\bf g}=\gamma_1[\mathbf h_0\times\mathbf n]/\varkappa$.
Note that the vectors $\mathbf w$ and ${\bf g}$ lie in the plane of the S film and  are perpendicular to the exchange field ${\bf h}_0$. The obtained functional allows the further calculation of the system critical temperature and depairing current.

\section{Anisotropy of critical temperature}\label{Sec_Tc}

The local enhancement of the critical temperature near the superconductor boundary described by the constant $\gamma_0$ gives rise to the surface superconductivity, i.e. formation of the localized superconducting state at temperatures above the critical temperature of the bulk superconductor. In this section we calculate the critical temperature of localized superconductivity (emerging above the bulk critical temperature $T_{c0}$) in finite-size superconducting film and demonstrate its dependence on the mutual orientation between the exchange field in the underlying ferromagnetic insulator with SOC and the the film boundary. 

To obtain the system critical temperature one can neglect the terms $\propto |\psi|^4$ in the free energy functional (\ref{Eq_2D}). Varying Eq.~(\ref{Eq_2D}) with respect to $\psi^*$ and considering the order parameter in the form $\psi(x,y)=\psi_0e^{(-s+ip)x+iqy}$ (which is the general solution of the linearized Ginzburg-Landau equation) with the certain amplitude $\psi_0$ and real constant parameters $q$, $p$ and $s$ (we require $s>0$ to guarantee the decay of the order parameter from the film edge) we obtain the following relation for the defined constants:
\begin{equation}\label{req}
-\alpha+\varkappa q^2-\varkappa(ip-s)^2-\gamma_0w_yq+i\gamma_0w_x(ip-s)=0.
\end{equation}
The corresponding free energy per unit length along the S film boundary as a function of the parameters reads:
\begin{equation}\label{F_res}
\begin{array}{c}{
F=\left|\psi_0\right|^2d\left\{-\gamma_0-\varkappa g_{x}p-\varkappa g_{y}q \right.}\\{}\\{
\left. +(2s)^{-1}\left[(-\alpha+\varkappa(s^2+p^2+q^2)-\gamma_0w_xp-\gamma_0w_yq\right]\right\}.}
\end{array}
\end{equation}
The transition from normal to superconducting state occurs when the free energy $F$ becomes negative for nonzero order parameter amplitude $\psi_0$. The maximal value of the constant $\alpha_m$ which enables $F=0$ for $\psi_0\neq 0$ defines the critical temperature $T_c$ of the corresponding phase transition:
\begin{equation}
T_c=T_{c0}(1+|\alpha_m|/\alpha_0).
\end{equation}
Using Eq.~(\ref{req}) one can express the parameters $s$ and $p$ via the constant $q$. Then solving the equation $F=0$ with the free energy defined by Eq.~(\ref{F_res}) we find the dependence $\alpha$ on $q$. Finally, maximizing the resulting function $\alpha(q)$ we obtain the desired maximal value $\alpha_m$ which determines the critical temperature:
\begin{equation}
\alpha_m=\frac{\gamma_0^2}{4\varkappa}\left\{
w_x^2+\left(2+g_xw_x\right)^2+\frac{\left[w_y+g_y\left(2+g_xw_x\right)\right]^2}{(1-g_{y}^2)}\right\}.
\end{equation}
In the absence of the SOC the problem is formally equivalent to the one previously analyzed in the context of the twinning plane superconductivity (see \cite{Buzdin-JETP-81}) with $\alpha_m=\gamma_0^2/\varkappa$. The presence of the uniform SOC which does not depend on $x$ (i.e. ${\bf g}=0$) the parameter $\alpha_m$ corresponding to the critical temperature takes the form $\alpha_m=\gamma_0^2(1+w^2/4)/\varkappa$ (where $w^2=w_x^2+w_y^2$) and does not depend on the orientation of the exchange field in the plane of the underlying ferromagnet. At the same time, the variation of the SOC constant near the boundary of the S film (${\bf g}\neq 0$) gives rise to the nontrivial dependence of $\alpha_m$ on the angle between the vector ${\bf w}$ and $y$ axis. In particular, we consider the situations when the exchange field is perpendicular (${\bf h}_0=h_0{\bf x_0}$, ${\bf w}=w{\bf y_0}$, ${\bf g}=g{\bf y_0}$) and parallel (${\bf h}_0=h_0{\bf y_0}$, ${\bf w}=w{\bf x_0}$, ${\bf g}=g{\bf x_0}$) to the film boundary. We denote the corresponding maximal values of the parameter $\alpha$ as $|\alpha_m|=\alpha_\perp$ and $|\alpha_m|=\alpha_\parallel$, respectively. The result has the form
\begin{align}
& \alpha_\perp=\frac{\gamma_0^2}{\varkappa}\left[
1+\frac{(w/2+g)^2}{1-g^2}\right],\label{Tc1}\\
& \alpha_\parallel=\frac{\gamma_0^2}{\varkappa}\left[
\left(1+\frac{gw}{2}\right)^2+\frac{w^2}{4}\right].\label{Tc2}
\end{align}
Eqs.~(\ref{Tc1})-(\ref{Tc2}) show that the critical temperature is higher in the case when the exchange field is perpendicular to the film edge which corresponds to the modulation of the order parameter phase along the film edge. Indeed, considering the limit $g\ll 1$ and expanding the denominator in Eq.~(\ref{Tc1}) into the Taylor series we find $\alpha_\perp-\alpha_\parallel=g^2\gamma_0^2/\varkappa+O\left(g^3\right)$.

Experimentally, the discovered difference in the critical temperature can be detected, e.g., by fabrication of a series of  superconducting stripes which have different orientation relative to the magnetization in the underlying ferromagnetic substrate and conventional critical temperature measurements.

\section{Anisotropy of critical current (diode effect)}\label{Sec_Ic}

In this section we analyze the diode effect and show that if the in-plane exchange field in the ferromagnetic insulator has a component perpendicular to the S film boundary the critical depairing current of the quasi-one-dimensional superconducting channel formed near this boundary above $T_{c0}$ depends on the current direction. 

The key ingredient required for the diode effect in superconducting systems with SOC is the impossibility to exclude the odd-degree terms over the Cooper pairs momentum by a certain phase transformation. Such impossibility may originate from the third-order gradients terms in the GL equation, non-uniform Meissner screening currents or, e.g., non-uniform SOC strength. In this paper we will focus on the latter situation assuming the SOC constant to be varying near the S film boundary ($\gamma_1\neq 0$). We also assume that the critical temperature is enhanced near the boundary $x=0$ which corresponds to $\gamma_0>0$.

Let us introduce the dimensionless units, namely, the dimensionless coordinates $\tilde x=x\gamma_0/\varkappa$ and $\tilde y=y\gamma_0/\varkappa$ as well as the order parameter $\psi=\gamma_0\varphi(\tilde x)\exp(i\tilde q\tilde y)/(\sqrt{\beta\varkappa})$ with the real amplitude $\varphi(\tilde x)$ and the wave vector $\tilde q=q\varkappa/\gamma_0$. For simplicity we assume that the exchange field has only the $x$ component  (${\bf h}_0=h_0{\bf x_0}$, ${\bf w}=w{\bf y_0}$, ${\bf g}=g{\bf y_0}$) so that the free energy density takes the form
\begin{equation}\label{f2}
\begin{array}{c}{\ds
f=\frac{\gamma_0^4}{\varkappa^2\beta}\left[- \varkappa\alpha\gamma_0^{-2}\varphi^2+\frac{1}{2}\varphi^4+\left(\frac{\partial\varphi}{\partial\tilde x}\right)^2\right.}\\{}\\{\ds \left. +(\tilde q^2-w\tilde q)\varphi^2-(1+ g\tilde q)\varphi^2\delta(\tilde x)\right].}
\end{array}
\end{equation}
\rev{The variational derivative of the corresponding free energy functional with respect to $\varphi$ gives the following equation for the order parameter:}
\begin{equation}\label{eq2}
-\varphi_{\tilde x\tilde x}- \varkappa\alpha\gamma_0^{-2}\varphi+\varphi^3+\left(\tilde q^2-w\tilde q\right)\varphi\rev{-(1+ g\tilde q)\varphi\delta(\tilde x)=0.}
\end{equation}
\rev{It is convenient, first, to solve this equation in the region $\tilde x>0$ where the delta-function is equal to zero so that 
\begin{equation}\label{eq2b}
-\varphi_{\tilde x\tilde x}- \varkappa\alpha\gamma_0^{-2}\varphi+\varphi^3+\left(\tilde q^2-w\tilde q\right)\varphi=0
\end{equation}
and then account for the last term of Eq.~(\ref{eq2}) by imposing the boundary condition}
\begin{equation}\label{eq22}
-\varphi_{\tilde x}(0)=(1+ g\tilde q)\varphi(0),
\end{equation}
\rev{which can be obtained directly from Eq.~(\ref{eq2}) by integrating it over the small vicinity of the point $\tilde x=0$.} \rev{The resulting equations (\ref{eq2b}--\ref{eq22}) are} very similar to the one previously studied in the context of the twinning plane superconductivity \cite{Averin-JETP-83, Khlustikov-1987, Khlustikov-UFN-88, Buzdin-JETP-81}. Following the calculation procedure used in Refs.~\cite{Averin-JETP-83}  we obtain the solution of Eq.~(\ref{eq2}) in the form
\begin{equation}
\varphi(\tilde x)=\frac{\sqrt{2\tau\left[\left(1+\tilde qg\right)^2-\tau\right]}}{\sqrt\tau\cosh(\tilde x\sqrt\tau)+(1+\tilde qg)\sinh(\tilde x\sqrt\tau)},
\end{equation}
where $\tau(\tilde q)=- \varkappa\alpha\gamma_0^{-2}-\tilde qw+\tilde q^2$. The critical temperature $T_c$ of the system is defined by Eq.~(\ref{Tc1}) and the resulting optimal values $\tilde q=q_0$ and $\tau=\tau_0$ corresponding to $T=T_c$ read
\begin{equation}\label{q0_def}
q_0=\frac{g+w/2}{1-g^2},\quad\tau_0=\frac{(2+gw)^2}{4(1-g^2)^2}.
\end{equation}

The density of the superconducting current corresponding to the certain wave vector $q$ is equal to a variational derivative $j=(c/\Phi_0)\partial f/\partial q$ and takes the form [see Eq.~(\ref{f2})]
\begin{equation}\label{Eq4}
j(\tilde x)=\frac{c\gamma_0^3\varphi^2(\tilde x)}{\Phi_0\varkappa\beta}\left[2\tilde q-w-g\delta(\tilde x)\right].
\end{equation}
Note that similarly to the result of \cite{Mironov-PRL-17} there is a current flowing inside the narrow region of the width $\sim\xi_0$ counterbalanced by the current flowing in the opposite direction far from the S film boundary. Substitution $\varphi(\tilde x)$ into Eq.~(\ref{Eq4}) and integration over the coordinate $x$ perpendicular to the film boundary gives the total current
\begin{equation}\label{curr2}
\begin{array}{c}{\ds
I=\frac{\varkappa d}{\gamma_0}\int_0^{+\infty}j(\tilde x)d\tilde x=}\\{}\\{\ds =\frac{2c\gamma_0^2 d}{\Phi_0\beta}(1+\tilde q g -\sqrt\tau)[2\tilde q-w-g(1+\tilde q g +\sqrt\tau)].}
\end{array}
\end{equation}
The factor $(1+q g -\sqrt\tau)$ in this expression is proportional to the Cooper pair density in the localized channel while the factor $[2q-w-g(1+q g +\sqrt\tau)]$ is proportional to the superconducting velocity.

\begin{figure}[t]
    \includegraphics[width=0.8\linewidth]{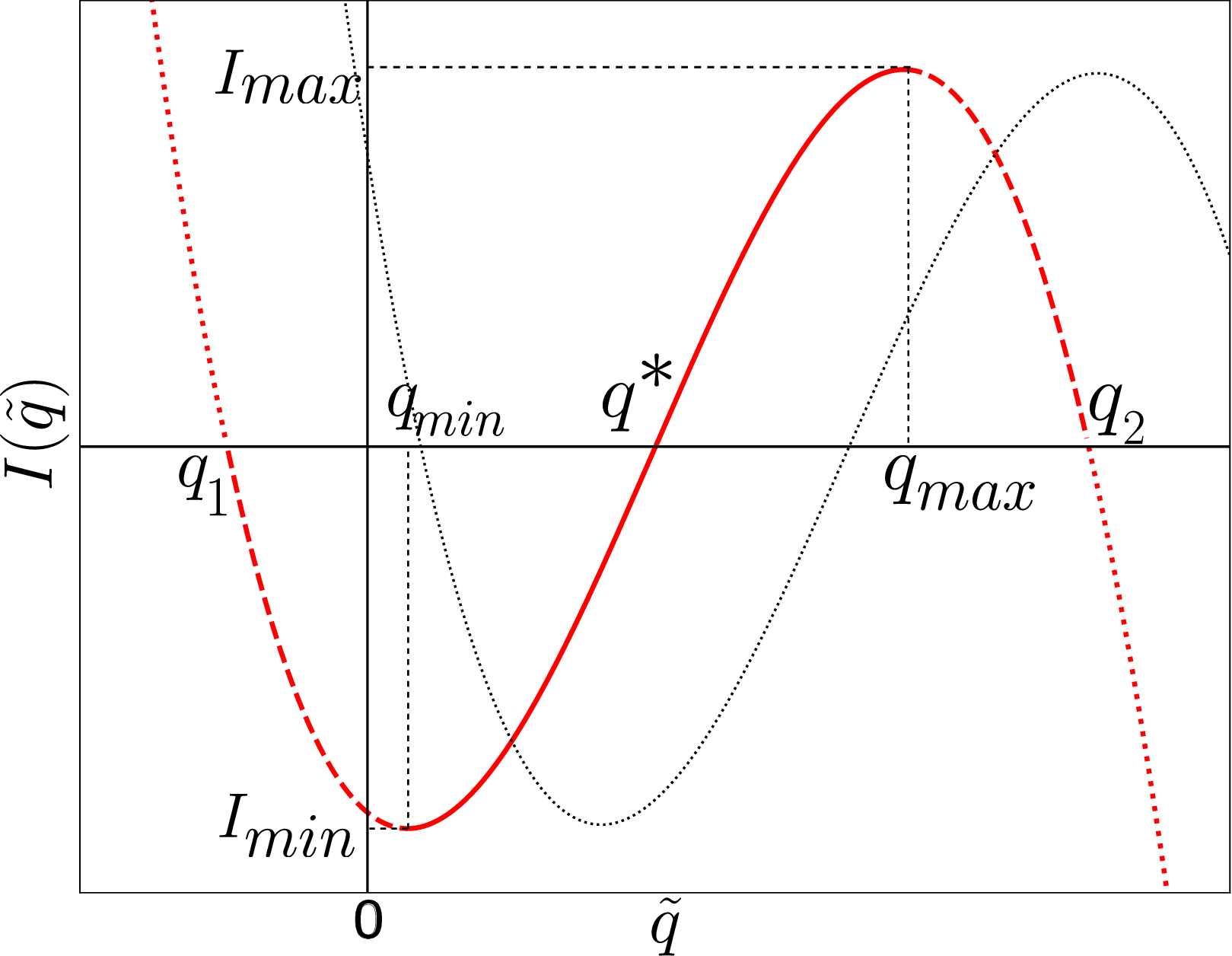}
    \caption{Typical dependence $I(\tilde q)$ for the values $\eta=0.05$, $ g =0.1$, $w=0.1$. The dotted line shows the $I(\tilde q)$ dependence for $\eta=0.05$, $g=0.1$, $w=0.3$.
    }
    \label{Fig_2}
\end{figure}

\begin{figure}[]
    \includegraphics[width=0.8\linewidth]{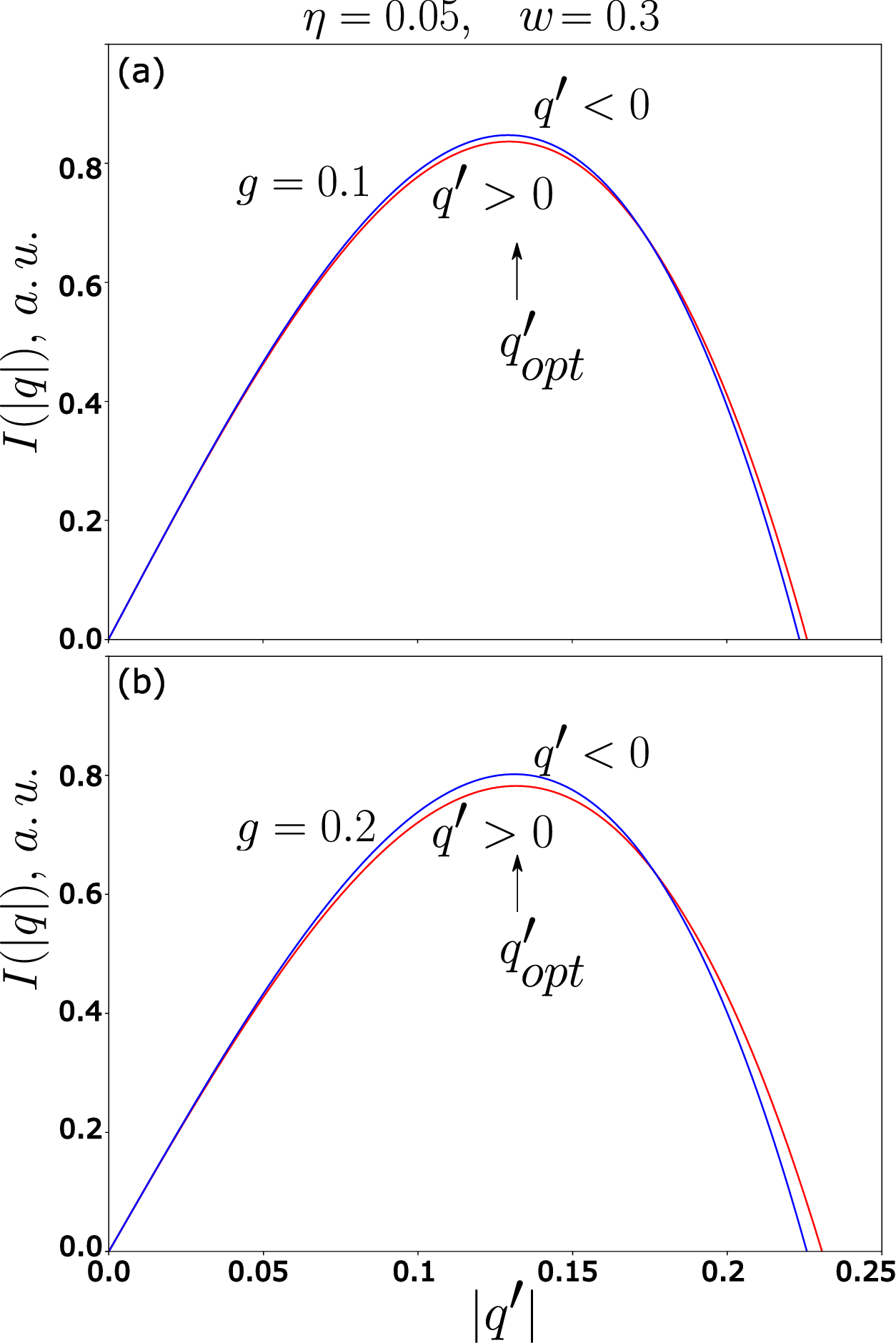}
    \caption{The typical dependencies $I(|q'|)$ illustrating the diode effect, i.e. the difference in the critical depairing current for two opposite current directions (compare blue and red curves). In our calculations we took $\eta=0.05$, $w=0.3$, and (a) $g=0.1$, (b) $g=0.2$.
    }
    \label{Fig_3}
\end{figure}

For the further analysis it is convenient to introduce the dimensionless parameter $\eta$ characterizing the system temperature $T$:
\begin{equation}
\eta=\frac{\alpha_0\varkappa}{\gamma_0^2}\frac{\left(T_c-T\right)}{T_{c0}}.
\end{equation}
Then the superconducting current becomes the function of the wave vector $\tilde q$ and the parameter $\eta$, i.e. $I=I(\tilde q,\eta)$ and the critical current $I_c$ at a fixed \rev{temperature} is defined as the maximal value of this function over $\tilde q$:
\begin{equation}
I_c(\eta)=\max_{\tilde q}I(\tilde q,\eta).
\end{equation}

The typical dependencies $I(\tilde q)$ are shown in Fig.~\ref{Fig_2}. These dependencies have two features: (i) the current is equal to zero at a finite value of phase gradient $\tilde q=q^\ast\neq 0$ and (ii) there is the asymmetry $I(q^\ast+q',\eta)\ne-I(q^\ast-q',\eta)$, where $q^\prime=\tilde q-q^\ast$. Exactly this asymmetry is responsible for the appearance of the diode effect. Note that the stable states for the fixed current are realized for the wave vectors in the range $q_{min}<\tilde q<q_{max}$ (the values $q_{min}$ and $q_{max}$ are defined in Fig.~\ref{Fig_2}) while the states with $\tilde q$ outside this range (i.a for $q_1<\tilde q<q_{min}$ and $q_{max}<\tilde q<q_2$) correspond to the local maximum of the system free energy. The ranges $\tilde q<q_1$ and $\tilde q>q_2$ formally correspond to $|\psi^2|<0$ and have no physical meaning.

\begin{figure}[hbt!]
	\includegraphics[width=0.8\linewidth]{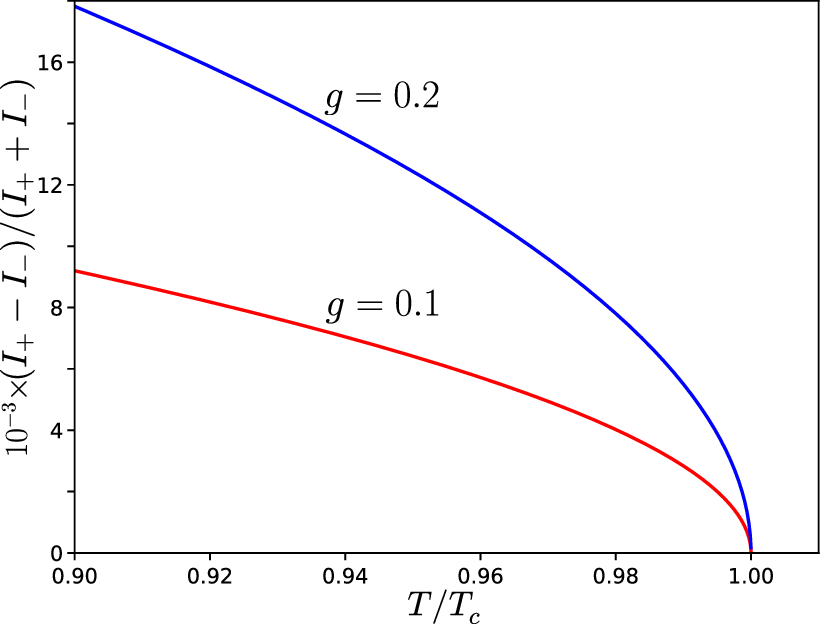}
	\caption{The dependence \rev{$(I_+-I_-)/(I_++I_-)$} on the structure temperature for the value $w=0.3$. The relative strength of the diode effect becomes stronger for the lower temperature.
	}
	\label{Fig_4}
\end{figure}

To demonstrate the diode effect arising in the system under consideration we calculate the values of the critical depairing current at a given temperature slightly below $T_c$ for two opposite current directions and show that they are different: 
\begin{equation}
\left|\min_{\tilde q}I(\tilde q,\eta)\right|\neq\max_{\tilde q}I(\tilde q,\eta).
\end{equation}
First, we calculate the wave vector $q^\ast(\eta)$ corresponding to the absence of the superconducting current in the localized channel at the fixed temperature characterized by the parameter $\eta$ [obviously, $q^\ast(0)=q_0$, see Eq.~(\ref{q0_def})]. In the limit $\eta\ll 1$ we get
\begin{equation}\label{qast}
q^\ast\approx q_0-\eta\frac{ g }{2(2+gw)}+O(\eta^2).
\end{equation}
The corresponding value of the parameter $\tau=\tau^\ast$ which defines the width of the superconducting channel reads
\begin{equation}\label{tauast}
\tau^\ast=\tau(q^\ast)\approx \tau_0-\eta\frac{2- g ^2}{2(1- g ^2)}+O(\eta^2).
\end{equation}
Substitution of Eqs.~(\ref{qast})-(\ref{tauast}) in the total current (\ref{curr2}) gives the following expression:
\begin{equation}
I=\frac{2c\gamma_0^2d}{\Phi_0\beta}q's\left(2\eta p-3q'\eta g sp^2-2q'^2sp+
5q'^3 g s^2p^2\right),
\end{equation}
where $s=1- g ^2$, $p=(2+ g h)^{-1}$, and inside the brackets we have neglected the terms of the order of $O(\eta^2,q'^4,\eta q'^2)$. The corresponding dependence $I(q^\prime)$ for two sets of parameters are shown in Fig.~\ref{Fig_3}. In the limit $\eta\ll 1$ the current reaches maximal value at $q^\prime=\sqrt{\eta s/3}$ so that the critical current $I_{\pm}$ for the two opposite directions reads
\begin{equation}\label{Ipm}
I_{\pm}=\frac{8c\gamma_0^2d}{9\Phi_0\beta}\left(\sqrt3\eta^{3/2}s^{3/2}p^{-2}\pm g \eta^2s^2p^2\right).
\end{equation}
In Eq.~(\ref{Ipm}) the first term stands for the usual Ginzburg-Landau expression for the critical current which is proportional to  $(T_c-T)^{3/2}$ while the second non-reciprocal contribution to the current is proportional to $(T_c-T)^2$ and describes the diode effect. As a result, the relative magnitude of the diode effect 
\begin{equation}\label{magn}
\frac{|I_+-I_-|}{I_++I_-}\propto\sqrt{T_c-T}.
\end{equation}
This temperature dependence qualitatively coincides with the one relevant to the 2D superconducting films if the higher-order gradients of the order parameter \cite{Daido-PRL-22,He-NJP-22} or a Meissner effect and consequent partial screening of the magnetic field \cite{Devizorova-PRB-21} are taken into account. Note also that the magnitude of the diode effect $|I_+-I_-|/|I_++I_-|$ (\ref{magn}) does not depend on the sign of the constant $\varepsilon_1$ in the initial free energy functional. The typical dependencies of the relative magnitude of the diode effect as a function of temperature are shown in Fig.~\ref{Fig_4}. 

\section{Conclusion}\label{Sec_Conc}

Thus, we have shown that local increase of the critical temperature and the subsequent variation of the SOC strength near the boundary of the superconducting film on top of ferromagnetic insulator gives rise to the formation of quasi-one-demensional superconducting channel with non-reciprocal transport properties. The critical temperature $T_c$ of such channel is shown to depend on the orientation of the exchange field in the ferromagnet relative to the superconducting film boundary. The maximum of $T_c$ is realized when the exchange field is perpendicular to the film boundary which corresponds to the emergence of the spontaneous currents flowing along the boundary. Moreover, for this orientation of the exchange field we have calculated the depairing critical current of the localized superconducting channel and showed that it depends on the current orientation which is a manifestation of the diode effect. The difference $\left|I_+-I_-\right|$ between two critical current values becomes stronger as the temperature $T$ decreases down from $T_c$, and the relative magnitude of the diode effect is $(I_+-I_-)/(I_++I_-)\propto\sqrt{T_c-T}$.

Note that the recent transport measurements performed for the elongated Al stripe on top of the YIG substrate \cite{Tikhomirov-JMMM-21} revealed the sensitivity of the superconducting transition temperature on the sample orientation relative to the magnetization in YIG. More precisely the critical temperature is higher for the perpendicular orientation of the magnetic moment to the strip boundary which is in perfect accordance with our prediction. Although such sensitivity may result from the difference in the penetration of the stray magnetic field to the superconducting stripe for two different sample orientation (which is very improbable due to the in-plane orientation of the YIG magnetic moment and much smaller size of the Al strip comparing to the YIG substrate), our calculations shows that the formation of the quasi-one-dimensional channels due to the local increase in the critical temperature may provide an alternative explanation. 

\rev{Note also that the non-reciprocal phenomena discussed in this paper are not specific to the case when the critical temperature and the strength of the SOC are modified near the sample edge. For example, similar phenomena should arise in superconducting film placed underneath the ferromagnetic film of the finite lateral size with the in-plane exchange field directed perpendicular to the film edge. In this case the effective exchange field acting on electron spins reveals a jump inside the S film which should give rise to the localized quasi-one-dimensional superconducting states with non-reciprocal transport properties. Also we expect the diode-type behavior for the S/F systems where the magnetic configuration inside the thick ferromagnet corresponds to the so-called closure domains (see, e.g., \cite{Hubert}), so that near the surface the domain walls separate the regions with the anti-parallel magnetic moments directed parallel to the ferromagnet surface and perpendicular to the domain wall. The S/F systems with such magnetic structure should support the regime of the domain-wall superconductivity with the one-dimensional superconducting channels emerging along the domain walls \cite{Houzet}. In this case the local increase in the critical temperature and, therefore, the described non-reciprocal phenomena can become much larger compared to the edge effect considered in our model provided the width of the domain-wall is slightly smaller than the superconducting coherence length and the exchange field in the ferromagnet is of the order of the bulk superconducting critical temperature. Note that the diode effect arising in the localized superconducting channel induced near the domain walls due to the orbital effect was theoretically considered in Ref.~\cite{Silaev_DW}.
}

In summary, our results show that the combination of the restricted geometry and spin-orbit coupling in superconductor/ferromagnet systems may lead to the new interesting effects with possible implementations in superconducting devices.

\section*{Acknowledgements}

The authors thank A. S. Mel'nikov and I. D. Tokman for fruitful discussions. This work was supported by the Russian Science Foundation (Grant No. 20-12-00053) in part related to the analysis of the diode effect. S. V. M. acknowledges the financial support of the Foundation for the Advancement of Theoretical Physics and Mathematics BASIS (Grant No. 23-1-2-32-1) in part related to the calculations of the critical temperature. The work of A. I. B. was supported by ANR SUPERFAST and the LIGHT S\&T Graduate Program and EU COST CA21144  Superqumap.

\end{document}